
%



\font\bigbf=cmbx10 scaled\magstep2

\font\twelverm=cmr10 scaled 1200    \font\twelvei=cmmi10 scaled 1200
\font\twelvesy=cmsy10 scaled 1200   \font\twelveex=cmex10 scaled 1200
\font\twelvebf=cmbx10 scaled 1200   \font\twelvesl=cmsl10 scaled 1200
\font\twelvett=cmtt10 scaled 1200   \font\twelveit=cmti10 scaled 1200

\skewchar\twelvei='177   \skewchar\twelvesy='60


\def\twelvepoint{\normalbaselineskip=12.4pt
  \abovedisplayskip 12.4pt plus 3pt minus 9pt
  \belowdisplayskip 12.4pt plus 3pt minus 9pt
  \abovedisplayshortskip 0pt plus 3pt
  \belowdisplayshortskip 7.2pt plus 3pt minus 4pt
  \smallskipamount=3.6pt plus1.2pt minus1.2pt
  \medskipamount=7.2pt plus2.4pt minus2.4pt
  \bigskipamount=14.4pt plus4.8pt minus4.8pt
  \def\rm{\fam0\twelverm}          \def\it{\fam\itfam\twelveit}%
  \def\sl{\fam\slfam\twelvesl}     \def\bf{\fam\bffam\twelvebf}%
  \def\mit{\fam 1}                 \def\cal{\fam 2}%
  \def\tt{\twelvett}
  \textfont0=\twelverm   \scriptfont0=\tenrm   \scriptscriptfont0=\sevenrm
  \textfont1=\twelvei    \scriptfont1=\teni    \scriptscriptfont1=\seveni
  \textfont2=\twelvesy   \scriptfont2=\tensy   \scriptscriptfont2=\sevensy
  \textfont3=\twelveex   \scriptfont3=\twelveex
 \scriptscriptfont3=\twelveex
  \textfont\itfam=\twelveit
  \textfont\slfam=\twelvesl
  \textfont\bffam=\twelvebf \scriptfont\bffam=\tenbf
  \scriptscriptfont\bffam=\sevenbf
  \normalbaselines\rm}



\def\beginlinemode{\endmode
  \begingroup\parskip=0pt
\obeylines\def\\{\par}\def\endmode{\par\endgroup}}
\def\beginparmode{\endmode
  \begingroup \def\endmode{\par\endgroup}}
\let\endmode=\par
{\obeylines\gdef\
{}}
\def\singlespace{\baselineskip=\normalbaselineskip}
\def\oneandathirdspace{\baselineskip=\normalbaselineskip
  \multiply\baselineskip by 4 \divide\baselineskip by 3}
\def\oneandahalfspace{\baselineskip=\normalbaselineskip
  \multiply\baselineskip by 3 \divide\baselineskip by 2}
\def\doublespace{\baselineskip=
\normalbaselineskip \multiply\baselineskip by 2}

\newcount\firstpageno
\firstpageno=1
\footline={\ifnum\pageno<\firstpageno{\hfil}%
\else{\hfil\twelverm\folio\hfil}\fi}
\let\rawfootnote=\footnote              
\def\footnote#1#2{{\rm\singlespace\parindent=0pt\rawfootnote{#1}{#2}}}
\def\raggedcenter{\leftskip=4em plus 12em \rightskip=\leftskip
  \parindent=0pt \parfillskip=0pt \spaceskip=.3333em \xspaceskip=.5em
  \pretolerance=9999 \tolerance=9999
  \hyphenpenalty=9999 \exhyphenpenalty=9999 }
\def\dateline{\rightline{\ifcase\month\or
  January\or February\or March\or April\or May\or June\or
  July\or August\or September\or October\or November\or December\fi
  \space\number\year}}
\def\received{\vskip 3pt plus 0.2fill
 \centerline{\sl (Received\space\ifcase\month\or
  January\or February\or March\or April\or May\or June\or
  July\or August\or September\or October\or November\or December\fi
  \qquad, \number\year)}}


\hsize=6.5truein
\vsize=8.9truein
\voffset=0.0truein
\parskip=\medskipamount
\twelvepoint            
\oneandathirdspace      
\overfullrule=0pt  



\def\title                      
  {\null\vskip 3pt plus 0.2fill
   \beginlinemode \doublespace \raggedcenter \bigbf}

\def\author                     
  {\vskip 3pt plus 0.2fill \beginlinemode
   \singlespace \raggedcenter}

\def\affil                      
  {\vskip 3pt plus 0.1fill \beginlinemode
   \oneandahalfspace \raggedcenter \sl}

\def\abstract                   
  {\vskip 3pt plus 0.3fill \beginparmode
   \oneandathirdspace\narrower}

\def\endtitlepage               
  {\endpage                     
   \body}

\def\body                       
  {\beginparmode}               

\def\subhead#1{                 
  \vskip 0.25truein             
  \noindent{{\it {#1}} \par}
   \nobreak\vskip 0.15truein\nobreak}

\def\refto#1{[#1]}           

\def\references              
  {\subhead{\bf References}         
   \beginparmode
   \frenchspacing \parindent=0pt \leftskip=1truecm
   \oneandathirdspace\parskip=8pt plus 3pt
 \everypar{\hangindent=\parindent}}

\gdef\refis#1{\indent\hbox to 0pt{\hss#1.~}}    

\gdef\journal#1, #2, #3, 1#4#5#6{               
    {\sl #1~}{\bf #2}, #3 (1#4#5#6)}           

\def\refstylenp{                
  \gdef\refto##1{ [##1]}                              
  \gdef\refis##1{\indent\hbox to 0pt{\hss##1)~}}      
  \gdef\journal##1, ##2, ##3, ##4 {                   
     {\sl ##1~}{\bf ##2~}(##3) ##4 }}

\def\refstyleprnp{              
  \gdef\refto##1{ [##1]}                              
  \gdef\refis##1{\indent\hbox to 0pt{\hss##1)~}}      
  \gdef\journal##1, ##2, ##3, 1##4##5##6{             
    {\sl ##1~}{\bf ##2~}(1##4##5##6) ##3}}

\def\prl{\journal Phys. Rev. Lett., }

\def\cmp{\journal Comm. Math. Phys., }

\def\ann{\journal Ann. Phys., }

\def\endreferences{\body}

\def\figurecaptions             
  { \beginparmode
   \subhead{Figure Captions}
}

\def\endpage                    
  {\vfill\eject}

\def\endpaper                   
  {\endmode\vfill\supereject}


\def\ref#1{Ref. #1}                     
\def\Ref#1{Ref. #1}                     

\def\frac#1#2{{\textstyle{#1 \over #2}}}

\def\sla{\raise.15ex\hbox{$/$}\kern-.57em}
\def\leaderfill{\leaders\hbox to 1em{\hss.\hss}\hfill}
\def\twiddle{\lower.9ex\rlap{$\kern-.1em\scriptstyle\sim$}}
\def\bigtwiddle{\lower1.ex\rlap{$\sim$}}
\def\gtwid{
\mathrel{\raise.3ex\hbox{$>$\kern-.75em\lower1ex\hbox{$\sim$}}}}
\def\ltwid{\mathrel{\raise.3ex\hbox
{$<$\kern-.75em\lower1ex\hbox{$\sim$}}}}
\def\square{\kern1pt\vbox{\hrule height 1.2pt\hbox
{\vrule width 1.2pt\hskip 3pt
   \vbox{\vskip 6pt}\hskip 3pt\vrule width 0.6pt}
\hrule height 0.6pt}\kern1pt}

\def\m@th{\mathsurround=0pt }
\def\leftrightarrowfill{$\m@th \mathord\leftarrow \mkern-6mu
 \cleaders\hbox{$\mkern-2mu \mathord- \mkern-2mu$}\hfill
 \mkern-6mu \mathord\rightarrow$}
\def\overleftrightarrow#1{\vbox{\ialign{##\crcr
     \leftrightarrowfill\crcr\noalign{\kern-1pt\nointerlineskip}
     $\hfil\displaystyle{#1}\hfil$\crcr}}}


\font\titlefont=cmr10 scaled\magstep3

\def\martinstyletitle                      
  {\null\vskip 3pt plus 0.2fill
   \beginlinemode \doublespace \raggedcenter \titlefont}

\font\twelvesc=cmcsc10 scaled 1200

\def\author                     
  {\vskip 3pt plus 0.2fill \beginlinemode
   \singlespace \raggedcenter\twelvesc}


\def\heading                            
  {\vskip 0.5truein plus 0.1truein      
\endheading
   \beginparmode \def\\{\par} \parskip=0pt \singlespace \raggedcenter}

\def\endheading
  {\par\nobreak\vskip 0.25truein\nobreak\beginparmode}

\def\subheading                         
  {\vskip 0.25truein plus 0.1truein
   \beginlinemode \singlespace \parskip=0pt \def\\{\par}\raggedcenter}

\def\tag#1$${\eqno(#1)$$}

\def\align#1$${\eqalign{#1}$$}

\def\aligntag#1$${\gdef\tag##1\\{&(##1)\cr}\eqalignno{#1\\}$$
  \gdef\tag##1$${\eqno(##1)$$}}

\def\endaligntag{}

\def\overset #1\to#2{{\mathop{#2}\limits^{#1}}}
\def\underset#1\to#2{{\let\next=#1\mathpalette\undersetpalette#2}}
\def\undersetpalette#1#2{\vtop{\baselineskip0pt
\ialign{$\mathsurround=0pt #1\hfil##\hfil$\crcr#2\crcr\next\crcr}}}


\def\ref#1{Ref.~#1}                     
\def\Ref#1{Ref.~#1}                     
\def\[#1]{[\cite{#1}]}
\def\cite#1{{#1}}
\def\(#1){(\call{#1})}
\def\call#1{{#1}}
\def\taghead#1{}
\def\frac#1#2{{#1 \over #2}}

\def\12{{1\over2}}

\def\sla{\raise.15ex\hbox{$/$}\kern-.57em}
\def\leaderfill{\leaders\hbox to 1em{\hss.\hss}\hfill}
\def\twiddle{\lower.9ex\rlap{$\kern-.1em\scriptstyle\sim$}}
\def\bigtwiddle{\lower1.ex\rlap{$\sim$}}
\def\gtwid{\mathrel{\raise.3ex\hbox{$>$
\kern-.75em\lower1ex\hbox{$\sim$}}}}
\def\ltwid{\mathrel{\raise.3ex\hbox{$<$
\kern-.75em\lower1ex\hbox{$\sim$}}}}
\def\square{\kern1pt\vbox{\hrule height 1.2pt\hbox
{\vrule width 1.2pt\hskip 3pt
   \vbox{\vskip 6pt}\hskip 3pt\vrule width 0.6pt}
\hrule height 0.6pt}\kern1pt}
\def\tdot#1{\mathord{\mathop{#1}\limits^{\kern2pt\ldots}}}

\def\pmb#1{\setbox0=\hbox{#1}%
  \kern-.025em\copy0\kern-\wd0
  \kern  .05em\copy0\kern-\wd0
  \kern-.025em\raise.0433em\box0 }
\def\qed{\vrule height 1.2ex width 0.5em}

\catcode`@=11
\newcount\tagnumber\tagnumber=0

\immediate\newwrite\eqnfile
\newif\if@qnfile\@qnfilefalse
\def\write@qn#1{}
\def\writenew@qn#1{}
\def\w@rnwrite#1{\write@qn{#1}\message{#1}}
\def\@rrwrite#1{\write@qn{#1}\errmessage{#1}}

\def\taghead#1{\gdef\t@ghead{#1}\global\tagnumber=0}
\def\t@ghead{}

\expandafter\def\csname @qnnum-3\endcsname
  {{\t@ghead\advance\tagnumber by -3\relax\number\tagnumber}}
\expandafter\def\csname @qnnum-2\endcsname
  {{\t@ghead\advance\tagnumber by -2\relax\number\tagnumber}}
\expandafter\def\csname @qnnum-1\endcsname
  {{\t@ghead\advance\tagnumber by -1\relax\number\tagnumber}}
\expandafter\def\csname @qnnum0\endcsname
  {\t@ghead\number\tagnumber}
\expandafter\def\csname @qnnum+1\endcsname
  {{\t@ghead\advance\tagnumber by 1\relax\number\tagnumber}}
\expandafter\def\csname @qnnum+2\endcsname
  {{\t@ghead\advance\tagnumber by 2\relax\number\tagnumber}}
\expandafter\def\csname @qnnum+3\endcsname
  {{\t@ghead\advance\tagnumber by 3\relax\number\tagnumber}}

\def\equationfile{%
  \@qnfiletrue\immediate\openout\eqnfile=\jobname.eqn%
  \def\write@qn##1{\if@qnfile\immediate\write\eqnfile{##1}\fi}
  \def\writenew@qn##1{\if@qnfile\immediate\write\eqnfile
    {\noexpand\tag{##1} = (\t@ghead\number\tagnumber)}\fi}
}

\def\callall#1{\xdef#1##1{#1{\noexpand\call{##1}}}}
\def\call#1{\each@rg\callr@nge{#1}}

\def\each@rg#1#2{{\let\thecsname=#1\expandafter\first@rg#2,\end,}}
\def\first@rg#1,{\thecsname{#1}\apply@rg}
\def\apply@rg#1,{\ifx\end#1\let\next=\relax%
\else,\thecsname{#1}\let\next=\apply@rg\fi\next}

\def\callr@nge#1{\calldor@nge#1-\end-}
\def\callr@ngeat#1\end-{#1}
\def\calldor@nge#1-#2-{\ifx\end#2\@qneatspace#1 %
  \else\calll@@p{#1}{#2}\callr@ngeat\fi}
\def\calll@@p#1#2{\ifnum#1>#2{\@rrwrite
{Equation range #1-#2\space is bad.}
\errhelp{If you call a series of equations by the notation M-N, then M and
N must be integers, and N must be greater than or equal to M.}}\else %
{\count0=#1\count1=
#2\advance\count1 by1\relax\expandafter\@qncall\the\count0,%
  \loop\advance\count0 by1\relax%
    \ifnum\count0<\count1,\expandafter\@qncall\the\count0,%
  \repeat}\fi}

\def\@qneatspace#1#2 {\@qncall#1#2,}
\def\@qncall#1,{\ifunc@lled{#1}{\def\next{#1}\ifx\next\empty\else
  \w@rnwrite{Equation number \noexpand\(>>#1<<)
has not been defined yet.}
  >>#1<<\fi}\else\csname @qnnum#1\endcsname\fi}

\let\eqnono=\eqno
\def\eqno(#1){\tag#1}
\def\tag#1$${\eqnono(\displayt@g#1 )$$}

\def\aligntag#1\endaligntag
  $${\gdef\tag##1\\{&(##1 )\cr}\eqalignno{#1\\}$$
  \gdef\tag##1$${\eqnono(\displayt@g##1 )$$}}

\def\eqalignno#1{\displ@y \tabskip\centering
  \halign to\displaywidth{\hfil$\displaystyle{##}$\tabskip\z@skip
    &$\displaystyle{{}##}$\hfil\tabskip\centering
    &\llap{$\displayt@gpar##$}\tabskip\z@skip\crcr
    #1\crcr}}

\def\displayt@gpar(#1){(\displayt@g#1 )}

\def\displayt@g#1 {\rm\ifunc@lled{#1}\global\advance\tagnumber by1
        {\def\next{#1}\ifx\next\empty\else\expandafter
        \xdef\csname
 @qnnum#1\endcsname{\t@ghead\number\tagnumber}\fi}%
  \writenew@qn{#1}\t@ghead\number\tagnumber\else
        {\edef\next{\t@ghead\number\tagnumber}%
        \expandafter\ifx\csname @qnnum#1\endcsname\next\else
        \w@rnwrite{Equation \noexpand\tag{#1} is
a duplicate number.}\fi}%
  \csname @qnnum#1\endcsname\fi}

\def\ifunc@lled#1{\expandafter\ifx\csname @qnnum#1\endcsname\relax}

\let\@qnend=\end\gdef\end{\if@qnfile
\immediate\write16{Equation numbers
written on []\jobname.EQN.}\fi\@qnend}

\catcode`@=12

\catcode`@=11
\newcount\r@fcount \r@fcount=0
\newcount\r@fcurr
\immediate\newwrite\reffile
\newif\ifr@ffile\r@ffilefalse
\def\w@rnwrite#1{\ifr@ffile\immediate\write\reffile{#1}\fi\message{#1}}

\def\writer@f#1>>{}
\def\referencefile{
  \r@ffiletrue\immediate\openout\reffile=\jobname.ref%
  \def\writer@f##1>>{\ifr@ffile\immediate\write\reffile%
    {\noexpand\refis{##1} = \csname r@fnum##1\endcsname = %
     \expandafter\expandafter\expandafter\strip@t\expandafter%
     \meaning\csname r@ftext
\csname r@fnum##1\endcsname\endcsname}\fi}%
  \def\strip@t##1>>{}}

\def\citeall#1{\xdef#1##1{#1{\noexpand\cite{##1}}}}
\def\cite#1{\each@rg\citer@nge{#1}}	

\def\each@rg#1#2{{\let\thecsname=#1\expandafter\first@rg#2,\end,}}
\def\first@rg#1,{\thecsname{#1}\apply@rg}	
\def\apply@rg#1,{\ifx\end#1\let\next=\relax
\else,\thecsname{#1}\let\next=\apply@rg\fi\next}

\def\citer@nge#1{\citedor@nge#1-\end-}	
\def\citer@ngeat#1\end-{#1}
\def\citedor@nge#1-#2-{\ifx\end#2\r@featspace#1 
  \else\citel@@p{#1}{#2}\citer@ngeat\fi}	
\def\citel@@p#1#2{\ifnum#1>#2{\errmessage{Reference range #1-
#2\space is bad.}%
    \errhelp{If you cite a series of references by the notation M-N, then M
and
    N must be integers, and N must be greater than or equal to M.}}\else%
 {\count0=#1\count1=#2\advance\count1
by1\relax\expandafter\r@fcite\the\count0,
  \loop\advance\count0 by1\relax
    \ifnum\count0<\count1,\expandafter\r@fcite\the\count0,%
  \repeat}\fi}

\def\r@featspace#1#2 {\r@fcite#1#2,}	
\def\r@fcite#1,{\ifuncit@d{#1}
    \newr@f{#1}%
    \expandafter\gdef\csname r@ftext\number\r@fcount\endcsname%
                     {\message{Reference #1 to be supplied.}%
                      \writer@f#1>>#1 to be supplied.\par}%
 \fi%
 \csname r@fnum#1\endcsname}
\def\ifuncit@d#1{\expandafter\ifx\csname r@fnum#1\endcsname\relax}%
\def\newr@f#1{\global\advance\r@fcount by1%
    \expandafter\xdef\csname r@fnum#1\endcsname{\number\r@fcount}}

\let\r@fis=\refis			
\def\refis#1#2#3\par{\ifuncit@d{#1}
   \newr@f{#1}%
   \w@rnwrite{Reference #1=\number\r@fcount\space is not cited up to
 now.}\fi%
  \expandafter
\gdef\csname r@ftext\csname r@fnum#1\endcsname\endcsname%
  {\writer@f#1>>#2#3\par}}

\def\ignoreuncited{
   \def\refis##1##2##3\par{\ifuncit@d{##1}%
    \else\expandafter\gdef
\csname r@ftext\csname r@fnum##1\endcsname\endcsname%
     {\writer@f##1>>##2##3\par}\fi}}

\def\r@ferr{\endreferences\errmessage{I was expecting to see
\noexpand\endreferences before now;  I have inserted it here.}}
\let\r@ferences=\references
\def\references{\r@ferences\def\endmode{\r@ferr\par\endgroup}}

\let\endr@ferences=\endreferences
\def\endreferences{\r@fcurr=0
  {\loop\ifnum\r@fcurr<\r@fcount
    \advance\r@fcurr by
1\relax\expandafter\r@fis\expandafter{\number\r@fcurr}%
    \csname r@ftext\number\r@fcurr\endcsname%
  \repeat}\gdef\r@ferr{}\endr@ferences}


\let\r@fend=\endpaper\gdef\endpaper{\ifr@ffile
\immediate\write16{Cross References written on
[]\jobname.REF.}\fi\r@fend}

\catcode`@=12

\citeall\refto		
\citeall\ref		%
\citeall\Ref		%

\ignoreuncited
\def\ss{\scriptscriptstyle}
\def\qed{\hfill\hbox{\vrule width 4pt height 6pt depth 1.5 pt}\medskip}
\def\u{{\Phi{}}}
\def\ubar{{\overline\Phi{}}}
\def\psibar{\overline\psi}
\def\chibar{\overline\chi}
\def\alphabar{\overline\alpha}
\def\betabar{\overline\beta}
\def\phibar{\overline\phi}
\def\proof{\par\noindent {\bf Proof:\ }}
\def\Q{{\cal Q}}
%
%
\pageno0
\line{\hfill July 1994}
\title NATURAL SYMMETRIES OF THE YANG-MILLS EQUATIONS
\author C. G. Torre
\affil Department of Physics
Utah State University
Logan, UT 84322-4415
USA
\abstract
\noindent{\bf Abstract:}

We define a natural generalized symmetry of the Yang-Mills equations as
an infinitesimal transformation of the Yang-Mills field, built in a local,
gauge invariant, and Poincar\'e invariant fashion from the Yang-Mills field
strength and its derivatives to any order, which maps solutions of the field
equations to other solutions.  On the jet bundle of Yang-Mills connections
we introduce a spinorial coordinate system that is adapted to the solution
subspace defined by the Yang-Mills equations.  In terms of this coordinate
system the complete classification of natural symmetries is carried out in a
straightforward manner.  We find that all natural symmetries of the
Yang-Mills equations stem from the gauge transformations admitted by the
equations.
\endtitlepage
\noindent{\bf 1. Introduction.}

Yang-Mills theory, by which we mean any non-Abelian gauge theory, has
provided a fruitful area of study for both physicists and mathematicians.
Physicists have used Yang-Mills theory to describe
the strong and electroweak interactions \refto{Cheng1984}.  Applications
of the
Yang-Mills equations in mathematics have been found in several areas; an
important
example is given by the recent discovery of an intimate relation between
reductions of the Yang-Mills equations and a large class of integrable
differential equations \refto{Ablowitz1993}.  Whether one is
interested in physical
or mathematical applications of the Yang-Mills equations, there are certain
basic structural properties of these equations that one would like to
understand.  One of the most fundamental properties to be examined is the
class of generalized symmetries admitted by the equations
\refto{Olver1993}, \refto{Bluman1989}.   Roughly speaking, by
generalized
symmetries we mean infinitesimal transformations of the fields that map
solutions to solutions.  The transformations are to be
constructed in a local fashion from the fields and their derivatives to any
finite order \refto{footnote1}.  Given a set of differential equations, the
presence
of symmetries is connected with the existence of conservation laws, the
construction of solution generating
techniques, and integrability properties of the equations \refto{Olver1993},
\refto{Bluman1989}, \refto{Fokas1987}, \refto{Mikhailov1991}.

There are, of course, manifest symmetries that are built into the
Yang-Mills equations, namely, the Poincar\' e and gauge symmetries.  The
Poincar\' e symmetry is responsible for ten first-order conservation laws,
while the
gauge symmetry leads to trivial conservation laws.  In recent years it has
been found that many non-linear differential equations admit ``hidden
symmetries''.  For example, the Sine-Gordon equation in 1+1 dimensions is
a non-linear wave equation with a built-in Poincar\' e symmetry group.
Remarkably, this equation admits an infinite number of higher-order
generalized symmetries and corresponding conservation laws
\refto{Olver1993}, and this fact is intimately associated with the
integrability of the Sine-Gordon equation.  In light of such examples, and
given the strong connection between the Yang-Mills equations and
integrable
systems, it is tempting to speculate that the Yang-Mills equations will admit
higher-order symmetries and conservation laws.

In this paper we begin a classification of all generalized symmetries
admitted by the
Yang-Mills equations on a flat four-dimensional spacetime.  Given the
manifest
gauge and Poincar\'e covariance of the Yang-Mills equations, it is
reasonable to search for symmetries that are constructed in a gauge and
Poincar\'e covariant manner from the Yang-Mills field strength and
its gauge-covariant derivatives.  We call such symmetries {\it
natural generalized symmetries}.  In order to
classify natural symmetries we borrow techniques from a recent
classification of all symmetries for the vacuum Einstein equations
\refto{IMA_CGT}.  The
principal tool used in \refto{IMA_CGT} was an adapted set of spinor
coordinates on the jet
space of Einstein metrics.  These coordinates derive, in part, from
Penrose's
notion of an ``exact set of fields'' \refto{Penrose1960},
\refto{Penrose1984}.  As noted by Penrose, an exact set of
fields exists for the Yang-Mills equations, and this leads, via a relatively
quick and straightforward analysis which is very similar to that of
\refto{IMA_CGT}, to a
complete classification of all natural symmetries of the Yang-Mills
equations.  Thus the power of combining spinor and jet space techniques
has a more general scope than merely in gravitation
theory.

In \S2 we summarize the preliminary results needed for our analysis.
The requirement that symmetries be built locally is handled by employing
the jet bundle description of Yang-Mills theory, and it is on the jet bundle
that
the adapted spinor coordinates are defined.  Various technical results
needed for our analysis are also presented.  In \S3 we analyze the linearized
Yang-Mills equations and classify the natural
symmetries.  We find that all natural symmetries of the Yang-Mills
equations stem from the gauge transformations admitted by the equations.
In \S4 we comment on the generalizations needed to effect a
complete classification of all symmetries of the Yang-Mills equations.
\taghead{2.}
\vskip0.25truein
\noindent{\bf 2. Preliminaries.}

We choose spacetime to be the manifold $M=R^4$ equipped with a flat
metric $\eta_{ab}$ of signature $(-+++)$.  The unique torsion-free
derivative operator compatible with $\eta_{ab}$ will be denoted by
$\partial_a$.  To define the Yang-Mills field we consider a principal
bundle $\pi:P\to M$ over spacetime with the structure group given by any
Lie group $G$.  Because every bundle over $R^4$ is trivial, we can
globally represent a connection on $\pi:P\to M$ by a 1-form $A_a$ on
$M$ taking values in the Lie algebra $g$ of $G$.  We call this 1-form the
{\it Yang-Mills field}.  The curvature of the connection is represented by a
2-form $F_{ab}$ on $M$ taking values in $g$, which will be called the
{\it Yang-Mills field strength}.  The field strength is given in terms of the
Yang-Mills field by
$$
F_{ab}=\partial_aA_b-\partial_bA_a + [A_a,A_b],\tag201
$$
where $[\cdot,\cdot]$ is the bracket of $g$.
If $\tau_\alpha$ is a basis for $g$, we write
$$
A_a=A_a^\alpha\tau_\alpha\qquad{\rm and}\qquad
F_{ab}=F_{ab}^\alpha\tau_\alpha.\tag202
$$
We then have
$$
F_{ab}^\alpha=\partial_aA^\alpha_b-\partial_bA^\alpha_a +
\kappa^\alpha_{\gamma\delta}A_a^\gamma A_b^\delta,\tag203
$$
where $\kappa^\alpha_{\gamma\delta}$ are the structure constants of the
Lie algebra $g$.

Given a representation $\rho$ of the group $G$ we have an associated
vector bundle $\pi\colon E\to M$.  The Yang-Mills field defines a
derivative operator $\nabla_a$ on sections $s\colon M\to E$ via
$$
\nabla_a s=\partial_a s + A_a\cdot s,\tag204
$$
where we use the raised dot ($\cdot$) to indicate the action of the Lie
algebra on sections that is defined by $\rho$.   The Yang-Mills field
strength measures the failure of this derivative operator to commute;  we
have the identity
$$
\nabla_{[a}\nabla_{b]} s={1\over2}F_{ab}\cdot s.\tag205
$$
The Yang-Mills field strength can be viewed as a (2-form-valued) section
of the vector bundle defined by the adjoint representation of $G$. We thus
have that
$$
\nabla_aF_{bc}=\partial_aF_{bc} + [A_a,F_{bc}],\tag206
$$
and the field strength satisfies the Bianchi identities
$$
\nabla_{[a}F_{bc]}=0.\tag207
$$

The {\it Yang-Mills field equations} are given by
$$
\nabla^a F_{ab}=\partial^aF_{ab} + [A^a,F_{ab}]=0.\tag208
$$
In terms of the basis $\tau_\alpha$ we have
$$
\nabla^a F_{ab}^\alpha=\partial^aF_{ab}^\alpha +
\kappa^\alpha_{\beta\gamma}A^{\beta a}F_{ab}^\gamma=0.\tag209
$$

Let $\pi\colon{\cal Q}\to M$ be the bundle of $g$-valued 1-forms on $M$.
A section $A\colon M\to \Q$ of this bundle is a Yang-Mills field $A_a(x)$.
Let $J^k({\cal Q})$ be the bundle of $k^{th}$-order jets of sections of
$\Q$ \refto{Olver1993}, \refto{Saunders1989}.  A point $\sigma$ in
$J^k(\Q)$ is defined by a spacetime point $x$, the Yang-Mills field at $x$
and all its derivatives to order $k$ at $x$.  A section $A\colon M\to\Q$
lifts to give a section $j^k(A)\colon M\to J^k({\cal Q})$, which is called the
$k$-jet of $A$.  If we write
$$
A_{a,b_1\cdots
b_k}(j^k(A)(x))=\partial_{b_1}\partial_{b_2}\cdots\partial_{b_k}
A_a(x),\tag210
$$
then a point $\sigma\in J^k(\Q)$ is given by
$$
\sigma=(x, A_a,A_{a,b_1}, \ldots,A_{a,b_1\cdots b_k}).\tag211
$$
The {\it total derivative} $D_cf$ of a function
$$
f=f(x, A_a,A_{a,b_1}, \ldots,A_{a,b_1\cdots b_k})\tag212
$$
on $J^k(\Q)$ is defined by
$$
D_cf={\partial f\over\partial x^c} + {\partial f\over\partial A_a^\alpha}
A^\alpha_{a,c} + {\partial f\over\partial A^\alpha_{a,b_1}}
A^\alpha_{a,cb_1} +\cdots+{\partial f\over\partial A^\alpha_{a,b_1\cdots
b_k}}A^\alpha_{a,cb_1\cdots b_k}.\tag213
$$
The main property of the total derivative is that it represents on the jet
bundle the effect of the derivative operator $\partial_a$ on fields.  More
precisely, if $f\colon J^{k-1}(\Q)\to R$ is a smooth function and $A\colon
M\to \Q$ is a Yang-Mills field with $k$-jet $j^k(A)\colon M\to J^k(\Q)$,
then we have the identity
$$
(D_a f)\circ j^k(A)(x)=\partial_a(f\circ j^{k-1}(A)(x)).\tag214
$$

The field equations \(208) define a submanifold
$$
{\cal R}^2\hookrightarrow J^2(\Q),
$$
which we call the {\it equation manifold}.  The $k^{th}$ (total) derivative
of the field equations defines the {\it prolonged equation manifold}
$$
{\cal R}^{k+2}\hookrightarrow J^{k+2}(\Q).
$$
A {\it generalized symmetry} for the field equations \(208) is an
infinitesimal map, depending locally on the independent variables, the
dependent variables, and the derivatives of the dependent variables to some
finite order, which carries solutions to nearby solutions.  Geometrically, a
generalized symmetry of order $k$ is a vector field on $J^k(\Q)$ which is
tangent to ${\cal R}^k$ and preserves the contact ideal associated to
$J^k(\Q)$ \refto{Saunders1989}.   A generalized symmetry of order $k$
for the Yang-Mills equations can be represented as a map from $J^k(\Q)$
into the bundle of $g$-valued 1-forms on $M$. We denote this map by
$C_a=C_a^\alpha\tau_\alpha$, and we write
$$
C_a = C_a(x, A_a,A_{a,b_1}, \ldots,A_{a,b_1\cdots b_k}).\tag2001
$$
We say a generalized symmetry is trivial if it vanishes on the prolonged
equation manifold.  Two generalized symmetries are deemed equivalent if
they differ by a trivial symmetry.  Any generalized symmetry of the form
\(2001) is equivalent to a generalized symmetry obtained by restricting
\(2001) to ${\cal R}^k$, that is, we can assume that $C_a$ is a map from
${\cal R}^k$ into the bundle of $g$-valued 1-forms on $M$.

The following proposition is easily established from the theory of
generalized symmetries \refto{Olver1993}.

\proclaim Proposition 2.1.
The functions
$$
C_a = C_a(x, A_a,A_{a,b_1}, \ldots,A_{a,b_1\cdots b_k}).
$$
represent a $k^{th}$-order generalized symmetry for the Yang-Mills field
equations if and only if
$$
\nabla^b\nabla_b C_a - \nabla^b\nabla_a C_b +
[C_b,F^b_{\phantom{b\,}a}]=0\qquad{\rm on}\ {\cal R}^{k+2},\tag216
$$
where
$$
\nabla_b C_a=D_b C_a + [A_b,C_a].\tag217
$$

\noindent Note that the defining equations \(216) for a generalized
symmetry are the linearized field equations.

Familiar examples of symmetries of the Yang-Mills equations stem from
the gauge and conformal invariance of these equations.  If $A_a^\alpha(x)$
is a solution to \(208), and $\phi\colon M\to M$ is a conformal isometry of
the spacetime $(M,\eta)$, then $\phi^* A_a^\alpha(x)$ is also a solution to
\(208).  Here we define $\phi^* A_a^\alpha(x)$ to be the pull-back of
$A_a^\alpha$ in which $A_a^\alpha(x)$ is viewed as a collection of
1-forms on $M$.   The infinitesimal form of this conformal symmetry
leads
to the following proposition.

\proclaim Proposition 2.2.
Let $\xi^a(x)$ be a conformal Killing vector field for the spacetime
$(M,\eta)$, then
$$
C_a=\xi^b(x) F_{ba}\tag218
$$
is a generalized symmetry of the Yang-Mills equations.

\noindent The infinitesimal transformation defined by \(218) is the ``gauge
covariant Lie derivative'' of $A_a$ along $\xi^b$.

Let $U\colon M\to P$ be a section of the principal bundle.  If $A_a$ is a
solution to the Yang-Mills equations \(208) then
$$
A^U_a=U^{-1}A_aU + U^{-1}\partial_aU\tag219
$$
is also a solution to \(208).  $A^U$ is called the {\it gauge transformation}
of $A$.  The infinitesimal form of the gauge transformations leads to the
following proposition.

\proclaim Proposition 2.3.
Let $\Lambda(x)=\Lambda^\alpha(x)\tau_\alpha$ be a $g$-valued function
on $M$, then
$$
C_a=\nabla_a \Lambda=\partial_a\Lambda(x) + [A_a,\Lambda(x)]\tag220
$$
is a generalized symmetry of the Yang-Mills equations.

The gauge symmetry of Proposition 2.3 can be generalized to the case
where $\Lambda$ is constructed locally from the Yang-Mills field and its
derivatives to any order.

\proclaim Proposition 2.4.
Let
$$
\Lambda=\Lambda(x, A_a,A_{a,b_1}, \ldots,A_{a,b_1\cdots b_{k-1}})
$$
be a $g$-valued function on $J^{k-1}(\Q)$, then
$$
C_a=\nabla_a \Lambda=D_a\Lambda + [A_a,\Lambda]\tag221
$$
is a $k^{th}$-order generalized symmetry of the Yang-Mills equations.

\noindent  We will call these symmetries {\it generalized gauge
symmetries}.

In this paper we will classify {\it natural} generalized symmetries.  These
are generalized symmetries that have a simple behavior under Poincar\'e
and gauge transformations of the Yang-Mills field.  More precisely, the
gauge transformations and isometries can be lifted (by prolongation
\refto{Olver1993}) to act on $J^k(\Q)$, and, in terms of these lifted
actions, we have the following definition of a natural generalized
symmetry.

\proclaim Definition 2.5.
Let $\phi\colon M\to M$ be an isometry of the spacetime $(M,\eta)$, and
$U\colon M\to P$ a section of the principal bundle.  A natural generalized
symmetry is a function
$$
C_a = C_a(x, A_a,A_{a,b_1}, \ldots,A_{a,b_1\cdots b_k})
$$
satisfying \(216), and such that for any $\phi$
$$
C_a(\phi^{-1}(x), \phi^*A_a,\phi^*A_{a,b_1}, \ldots,\phi^*A_{a,b_1\cdots
b_k})=\phi^*C_a(x, A_a,A_{a,b_1}, \ldots,A_{a,b_1\cdots b_k}),\tag222
$$
and for any $U$
$$
C_a(x, A^U_a,A^U_{a,b_1}, \ldots,A^U_{a,b_1\cdots b_k})=U^{-
1}C_aU(x, A_a,A_{a,b_1}, \ldots,A_{a,b_1\cdots b_k}).\tag223
$$

We remark that, according to this definition, a generalized gauge
symmetry can be a natural generalized symmetry, but the conformal
symmetry of Proposition 2.2 is {\it not} a natural symmetry.  We also note
that we could have defined a natural symmetry using the full conformal
group; by only using the Poincar\'e subgroup we put fewer restrictions on
the allowed symmetries.

To elucidate the structure of a natural generalized symmetry we will
construct a set of adapted coordinates for $J^k(\Q)$.  To this end, let us
define
$$
A^\alpha_{b_0b_1\cdots
b_k}=\partial_{(b_1}\cdots\partial_{b_k}A^\alpha_{b_0)},\quad
k=0,1,\ldots\tag224
$$
and
$$
Q^\alpha_{b_0,b_1\cdots b_k}=\nabla_{(b_k}\nabla_{b_{k-
1}}\cdots\nabla_{b_{2}}F^\alpha_{b_1)b_0},
\quad k=1,2,\ldots.\tag225
$$
Both $A^\alpha_{b_0b_1\cdots b_k}$ and $Q^\alpha_{b_0,b_1\cdots
b_k}$ depend on the Yang-Mills field and its first $k$ derivatives; we
denote these variables by $A^k$ and $Q^k$.  Each of these variables is
algebraically irreducible in the sense that
$$
A^\alpha_{b_0b_1\cdots b_k}=A^\alpha_{(b_0b_1\cdots b_k)},\tag226
$$
$$
Q^\alpha_{b_0,b_1\cdots b_k}=Q^\alpha_{b_0,(b_1\cdots
b_k)}\qquad{\rm and}\qquad Q^\alpha_{(b_0,b_1\cdots b_k)}=0.\tag227
$$
We have the identity
$$
\partial_{b_1}\cdots\partial_{b_k}A^\alpha_{b_0}
=A^\alpha_{b_0b_1\cdots b_k}+{k\over k+1}Q^\alpha_{b_0,b_1\cdots
b_k}
+L^\alpha_{b_0b_1\cdots b_k},\tag228
$$
where $L^\alpha_{b_0b_1\cdots b_k}$ depends on the Yang-Mills field
and its derivatives to order $k-1$.  From this identity it is straightforward
to show that coordinates for $J^k(\Q)$ are given by
$$
(x,A^\alpha_{b_0},A^\alpha_{b_0b_1},Q^\alpha_{b_0,b_1},\ldots,
A^\alpha_{b_0b_1\cdots b_k},Q^\alpha_{b_0,b_1\cdots b_k}).\tag229
$$
Here we have taken the convenient liberty of using the same symbols $Q$
and $A$ to denote the fields on spacetime and functions on jet space.
Every generalized symmetry can be expressed as a function of the variables
\(229):
$$
C^\alpha_a=C^\alpha_a(x,A^\alpha_{b_0},A^\alpha_{b_0b_1},Q^\alpha_{b
_0,b_1},\ldots,A^\alpha_{b_0b_1\cdots b_k},Q^\alpha_{b_0,b_1\cdots
b_k}).\tag230
$$

We can now characterize natural generalized symmetries as follows.

\proclaim Proposition 2.6.
Let $C_a$ be a natural generalized symmetry of the Yang-Mills
equations of order $k$.  Then $C_a$ can be expressed as a function
of the variables $Q^\alpha_{b_0,b_1\cdots b_l}$ for $l=1,2,\ldots,k$, that
is,
$$
C_a=C_a(Q^\alpha_{b_0,b_1},
Q^\alpha_{b_0,b_1b_2},\ldots,Q^\alpha_{b_0,b_1\cdots b_k}).\tag231
$$

\proof
We begin by analyzing the requirement \(223).  Let $C^\alpha_a$ be given
as in \(230).  Let $U(t)\colon R\times M\to P$ be a 1-parameter family of
gauge transformations such that $U(0)$ is the identity transformation.  The
derivative
$$
\Lambda={dU\over dt}\big|_{t=0}\tag232
$$
is a $g$-valued function on $M$ defining an infinitesimal gauge
transformation.  Under an infinitesimal transformation $\Lambda$
associated to $U(t)$ we have that
$$
 {d\over dt}A^{U(t)\alpha}_{b_1\cdots b_k}\big|_{t=0}
=\Lambda^\alpha_{,b_1\cdots b_{k}} + \{\star\},\tag233
$$
where $\Lambda^\alpha_{,b_1\cdots b_l}$, for $l=0,1,2,\ldots,k$ (along
with $x$) defines the $k$-jet of $\Lambda$, and $\{\star\}$ denotes terms
involving $A^l$ and $\Lambda^\alpha_{,b_1\cdots b_l}$, for
$l=0,1,2,\ldots,k-1$.  We also have that
$$
{d\over dt}Q^{U(t)\alpha}_{a,b_1\cdots b_k}\big|_{t=0}
=\kappa_{\gamma\beta}^\alpha\Lambda^\beta Q^\gamma_{a,b_1\cdots
b_k}.\tag234
$$
We now demand that \(223) holds for any $U(t)$ and differentiate this
equation with respect to $t$ to find
$$
{\partial C^\alpha_a\over\partial A^\beta_{b_0\cdots b_{k}}}
\Lambda^\beta_{,b_0\cdots b_{k}} + R^\alpha_a
=\Lambda^\gamma\kappa_{\beta\gamma}^\alpha C^\beta_a,\tag235
$$
where $R^\alpha_a$ is independent of the variables
$\Lambda^\beta_{,b_0\cdots b_{k}}$.  Equation \(235) must hold for all
values of $\Lambda^\beta_{,b_0\cdots b_{k}}$ and this implies that
$$
{\partial C^\alpha_a\over\partial A^\beta_{b_0\cdots b_{k}}}=0.\tag236
$$
A simple induction argument then establishes that
$$
{\partial C^\alpha_a\over\partial A^\beta_{b_0\cdots b_{l}}}=0\tag237
$$
for $l=0,1,\ldots,k$.  Thus we have
$$
C^\alpha_a=C^\alpha_a(x,
Q^\alpha_{b_0,b_1},Q^\alpha_{b_0,b_1b_2},\ldots,Q^\alpha_{b_0,b_1
\cdots b_k}).\tag238
$$

It remains to be shown that $C^\alpha_a$ is independent of $x$.  Let
$x^\mu$ be a global inertial coordinate chart on $M$, and let $\xi^a$ be a
translational Killing vector field.  In the chart $x^\mu$ the components
$\xi^\mu$ are any set of constants:
$$
{\partial\xi^\mu\over\partial x^\nu}=0.\tag239
$$
If we demand that \(222) be satisfied for all translational isometries we
have that
$$
C^\alpha_a(x^\mu-\xi^\mu,Q^\alpha_{b_0,b_1},
Q^\alpha_{b_0,b_1b_2},\ldots,Q^\alpha_{b_0,b_1\cdots b_k})
=C^\alpha_a(x^\mu,Q^\alpha_{b_0,b_1},
Q^\alpha_{b_0,b_1b_2},\ldots,Q^\alpha_{b_0,b_1\cdots b_k})\tag240
$$
for any constants $\xi^\mu$,
which implies that
$$
C^\alpha_a=C^\alpha_a(Q^\alpha_{b_0,b_1},
Q^\alpha_{b_0,b_1b_2},\ldots,Q^\alpha_{b_0,b_1\cdots b_k}).\tag241
$$
\qed

If $C_a=C_a(Q^\alpha_{b_0,b_1},
Q^\alpha_{b_0,b_1b_2},\ldots,Q^\alpha_{b_0,b_1\cdots b_k})$ is a natural
generalized symmetry of the Yang-Mills equations, then it must satisfy the
linearized equations \(216) at each point of ${\cal R}^{k+2}$.  To classify
solutions to the linearized equations we will construct an explicit
parametrization of the prolonged equation manifolds.

In the following proposition $[Q^\alpha_{b_0,b_1\cdots b_k}]_{\ss\rm
tracefree}$
denotes the completely trace-free part of the tensor
$Q^\alpha_{b_0,b_1\cdots b_k}$ with respect to the metric $\eta_{ab}$.

\proclaim Proposition 2.7.
The variables
$$
(x,A^\alpha_{b_0},A^\alpha_{b_0b_1},\ldots,A^\alpha_{b_0b_1\cdots
b_k},[Q^\alpha_{b_0,b_1}]_{\ss\rm tracefree}, \ldots,
[Q^\alpha_{b_0,b_1\cdots b_k}]_{\ss\rm tracefree})\tag242
$$
form a global coordinate system for ${\cal R}^k$.

\proof
The prolonged equation manifold ${\cal R}^k$ can be defined by $k$-jets
which satisfy
$$
\eta^{mn}\nabla_{(b_1}\cdots \nabla_{b_l)}\nabla_m F_{na}=0\tag243
$$
for $l=0,1,2,\ldots,k-2$.
We express these equations in terms of the variables $Q^j$ via the identity
$$
\eta^{mn}\nabla_{(b_1}\cdots \nabla_{b_l)}\nabla_m F_{na}
=\left({l+2\over l+3}\right)\eta^{mn}[Q_{a,mnb_1\cdots b_l}-
Q_{n,amb_1\cdots b_l}]
+L_{ab_1\cdots b_l},\tag244
$$
where $L_{ab_1\cdots b_l}=L_{a(b_1\cdots b_l)}$ depends on $Q^j$ for
$j=1,2,\ldots,l$.  From \(244) we have that
$$
\eta^{ab_1}L_{ab_1\cdots b_l}=0\quad{\rm on}\ {\cal R}^l.\tag245
$$

Let $S^p$ denote the vector space of tensors with the algebraic symmetries
\(227) of $Q^p$.  Denote by $S_0^p$ the subspace of totally trace-free
tensors.  Let $T^p$ be the vector space of tensors with the symmetries of
$L_{ab_1\cdots b_p}$ and satisfying the trace condition \(245).  Define a
linear map $\Psi^p\colon S^{p+2}\to T^p$ which takes $W_{a,mnb_1\cdots
b_p}\in S^{p+2}$ into $V_{ab_1\cdots b_p}\in T^p$ by the rule
$$
V_{ab_1\cdots b_p}=\left({l+2\over
l+3}\right)\eta^{mn}[W_{a,mnb_1\cdots b_p}-W_{n,amb_1\cdots
b_p}].\tag246
$$
It is straightforward to show that
$$
{\rm Ker}\ \Psi^p=S_0^{p+2},\tag247
$$
and
$$
{\rm Im}\ \Psi^p=T^p.\tag248
$$

By virtue of \(247) and \(248), each point in ${\cal R}^k, k=2,3,\ldots$,
can be
uniquely determined as follows.  Let us begin with ${\cal R}^2$.  Choose
$x$, $A_a$ and $Q_{a,b_1}$, $A_{ab_1}$, and $A_{ab_1b_2}$
arbitrarily.  In equation \(244) with $l=0$ we have $L_a=0$, and so, from
\(247) and \(248), we solve \(243) by setting all traces of $Q^2$ to
zero.  ${\cal R}^2$ is thus parametrized by
$$
(x, A_a, A_{ab_1}, Q_{a,b_1}, A_{ab_1b_2}, [Q_{a,b_1b_2}]_{\rm \ss
tracefree}).\tag249
$$
Now we consider ${\cal R}^3$.  We choose the coordinates \(249) and
$A^3$ arbitrarily.  In the identity \(244) for $l=1$ we have that
$L_{ab_1}$ depends on $[Q^1]_{\ss\rm tracefree}$ only.  By virtue of the
surjectivity \(248) of the map $\Psi^1$ we can solve \(243).  By virtue of
\(247) the solution will be uniquely parametrized by $[Q^3]_{\rm \ss
tracefree}$, $A^3$, and the variables \(249).  By iterating this procedure,
we can build {\it every} solution to \(243), which is viewed as an equation
on $J^k(\Q)$, and the solutions will be {\it uniquely} parametrized by the
variables \(242).\qed

In principle, the variables \(242) can be used to analyze the linearized
equation \(216), but the resulting equations are still rather complicated.
Considerable simplifications can be obtained by using a spinor
representation
of the variables $[Q^k]_{\ss \rm tracefree}$.  Hence we now describe a
spinorial coordinate system on ${\cal R}^k$.  We remark that while all of
the results presented to this point are essentially independent of the
spacetime dimension, our use of spinors will limit the validity of
subsequent results to a 4-dimensional spacetime.

We begin with a brief summary of notation; for more details on spinors,
see \refto{Penrose1984}.  The spacetime metric and associated derivative
operator have the spinor representation
$$
\eta_{ab}\longleftrightarrow \epsilon_{\ss AB}\epsilon_{\ss A^\prime
B^\prime}
$$
and
$$
\partial_a\longleftrightarrow \partial_{\ss AA^\prime}.
$$
The $\epsilon$ spinors are skew-symmetric and non-degenerate at each
point of $M$.
The Yang-Mills field and field strength have the spinor representation
$$
A_b\longleftrightarrow A_{\ss BB^\prime}
$$
and
$$
F_{ab}\longleftrightarrow F_{\ss AA^\prime BB^\prime}
=\u_{\ss AB}\epsilon_{\ss A^\prime B^\prime}
+\ubar_{\ss A^\prime B^\prime }\epsilon_{\ss A B}.\tag250
$$
In \(250) the $g$-valued spinor fields $\u$ and $\ubar$ are symmetric,
$$
\u_{\ss AB}=\u_{\ss (AB)}\qquad {\rm and}\qquad\ubar_{\ss A^\prime
B^\prime }=\ubar_{\ss (A^\prime B^\prime) },\tag251
$$
and correspond to the self-dual and anti-self-dual part of the field strength.

The Bianchi identities \(207) take the spinor form
$$
\nabla^{\ss B}_{\ss A^\prime}\u_{\ss AB}
=\nabla^{\ss B^\prime}_{\ss A}\ubar_{\ss A^\prime B^\prime},\tag252
$$
while the identities \(205) become
$$
\nabla_{\ss X^\prime(A}\nabla_{\ss B)}^{\ss X^\prime} v
=-i\u_{\ss AB}\cdot v\tag253
$$
and
$$
\nabla_{\ss X(A^\prime}\nabla_{\ss B^\prime)}^{\ss X} v
=-i\ubar_{\ss A^\prime B^\prime }\cdot v.\tag254
$$
Given the identities \(252), the field equations \(208) are equivalent to
$$
\nabla_{\ss A^\prime}^{\ss B}\u_{\ss AB}=0
=\nabla_{\ss A}^{\ss B^\prime}\ubar_{\ss A^\prime B^\prime}.\tag255
$$

We now present a spinor representation of $[Q^k]_{\rm\ss tracefree}$.

\proclaim Proposition 2.8.
Let the $g$-valued tensor $Q^k$ be defined as in \(225), and let
$[Q^k]_{\rm\ss tracefree}$ have the spinor representation
$$
[Q_{b_0,b_1\cdots b_k}]_{\ss\rm tracefree}\longleftrightarrow
Q^{\ss B_0^\prime, B_1^\prime\cdots B_k^\prime}_{\ss B_0,B_1\cdots
B_k},
$$
then $[Q^k]_{\rm\ss tracefree}$ admits the unique spinor decomposition
$$
Q^{\ss B_0^\prime, B_1^\prime\cdots B_k^\prime}_{\ss B_0,B_1\cdots
B_k}
=\epsilon^{\ss B_0^\prime(B_1^\prime}\u^{\ss B_2^\prime\cdots
B_k^\prime)}_{\ss B_0B_1\cdots B_k}
+\epsilon_{\ss B_0(B_1}\ubar^{\ss B_0^\prime B_1^\prime\cdots
B_k^\prime}_{\ss B_2\cdots B_k)},\tag266
$$
where $\u^{\ss B_1^\prime\cdots B_{k-1}^\prime}_{\ss B_1\cdots
B_{k+1}}$ and $\ubar^{\ss B_1^\prime\cdots B_{k+1}^\prime}_{\ss
B_1\cdots B_{k-1}}$ are the totally symmetric spinors
$$
\u^{\ss B_1^\prime\cdots B_{k-1}^\prime}_{\ss B_1\cdots B_{k+1}}
=\nabla_{\ss (B_1}^{\ss (B_1^\prime}\cdots
\nabla_{\ss B_{k-1}}^{\ss B^\prime_{k-1})}\u_{\ss B_kB_{k+1})}\tag267
$$
$$
\ubar^{\ss B_0^\prime\cdots B_{k+1}^\prime}_{\ss B_0\cdots B_{k-1}}
=\nabla_{\ss (B_0}^{\ss (B_0^\prime}\cdots
\nabla_{\ss B_{k-1})}^{\ss B^\prime_{k-1}}\ubar^{\ss B_k^\prime
B^\prime_{k+1})}.\tag268
$$

\proof
{}From the first symmetry given in \(227) and the trace-free requirement on
the indices $b_1\cdots b_k$, it is readily shown that
$$
Q^{\ss B_0^\prime, B_1^\prime\cdots B_k^\prime}_{\ss B_0,B_1\cdots
B_k}
=Q^{\ss B_0^\prime, (B_1^\prime\cdots B_k^\prime)}_{\ss B_0,B_1\cdots
B_k}
=Q^{\ss B_0^\prime ,B_1^\prime\cdots B_k^\prime}_{\ss B_0,(B_1\cdots
B_k)}.\tag269
$$
The requirement
$$
\eta^{b_0b_1}Q_{b_0,b_1\cdots b_k}=0\tag270
$$
and the cyclic symmetry in \(227) leads to the algebraic form \(266).
In \(266) the spinors $\u^{\ss B_2^\prime\cdots B_{k}^\prime}_{\ss
B_0\cdots B_{k}}$ and $\ubar^{\ss B_0^\prime\cdots B_{k}^\prime}_{\ss
B_2\cdots B_{k}}$ are uniquely defined by
$$
\u^{\ss B_2^\prime\cdots B_k^\prime}_{\ss B_0B_1\cdots B_k}
={k\over k+1}\epsilon_{\ss B_0^\prime B_1^\prime}
Q^{\ss B_0^\prime, B_1^\prime\cdots B_k^\prime}_{\ss B_0,B_1\cdots
B_k}\tag271
$$
and
$$
\ubar^{\ss B_0^\prime B_1^\prime\cdots B_k^\prime}_{\ss B_2\cdots
B_k}={k\over k+1}\epsilon^{\ss B_0 B_1}
Q^{\ss B_0^\prime, B_1^\prime\cdots B_k^\prime}_{\ss B_0,B_1\cdots
B_k}.\tag272
$$
We use the decomposition \(250) in the spinor representation of \(225),
then, using \(266), \(271), and \(272), we can solve for $\u^{\ss
B_2^\prime\cdots B_{k}^\prime}_{\ss B_0\cdots B_{k}}$ and $\ubar^{\ss
B_0^\prime\cdots B_{k}^\prime}_{\ss B_2\cdots B_{k}}$ to find \(267)
and \(268).\qed

{}From Propositions 2.7 and 2.8 we can now define a spinorial coordinate
system on ${\cal R}^k$.

\proclaim Proposition 2.9.
The variables
$$
(x,A_{b_0},A_{b_0b_1},\ldots,A_{b_0\cdots b_k}, \u_{\ss
B_0B_1},\ubar_{\ss B_0^\prime B_1^\prime},\ldots,\u_{\ss B_0\cdots
B_k}^{\ss B_2^\prime\cdots B_{k}^\prime},\ubar^{\ss B_0^\prime \cdots
B_k^\prime}_{\ss B_2\cdots B_{k}})\tag273
$$
define a global coordinate chart on ${\cal R}^k$.

\noindent We remark that to pass between the coordinates \(242) and \(273)
we use any soldering form $\sigma_a^{\ss AA^\prime}$ such that
$$
\eta_{ab} = \sigma_a^{\ss AA^\prime}\sigma_{b\ss AA^\prime}.\tag274
$$

The spinor variables $\u^{\ss B_1^\prime\cdots B_{k-1}^\prime}_{\ss
B_1\cdots B_{k+1}}$ and $\ubar^{\ss B_1^\prime\cdots
B_{k+1}^\prime}_{\ss B_1\cdots B_{k-1}}$ will play a fundamental role
in our symmetry analysis.  Their role as coordinates for ${\cal R}^k$
stems from the fact that $\u_{\ss AB}$ and $\ubar_{\ss A^\prime
B^\prime}$ form what Penrose calls an ``exact set of fields'' for the
Yang-Mills equations \refto{Penrose1984}.  Henceforth we will call the
fields \(267) and \(268) the {\it Penrose fields} and denote them by $\u^k$
and $\ubar^k$.

By virtue of the identities \(252), \(253), \(254), and equations \(255), the
Penrose fields satisfy the following {\it structure equation} on the
prolonged equation manifolds.  See \refto{Penrose1984} for details.
\proclaim Proposition 2.10.
The spinorial covariant derivative of $\u^{\ss B_1^\prime\cdots B_{k-
1}^\prime}_{\ss B_1\cdots B_{k+1}}$, when evaluated on ${\cal R}^k$, is
given by
$$
\nabla_{\ss A}^{\ss A^\prime}\u^{\ss B_1^\prime\cdots B_{k-
1}^\prime}_{\ss B_1\cdots B_{k+1}}
=\u^{\ss A^\prime B_1^\prime\cdots B_{k-1}^\prime}_{\ss AB_1\cdots
B_{k+1}}
+\{\star\},\tag275
$$
where $\{\star\}$ denotes a spinor (and $g$) -valued function of the
Penrose fields $\u^1$, $\ubar^1$, \dots, $\u^{k-1}$, $\ubar^{k-1}$.

\noindent An analogous result holds for the complex conjugate Penrose
fields $\ubar^k$.  Proposition 2.10 is central to our generalized symmetry
analysis.

{}From Propositions 2.6--2.9 we now have the following restriction on the
domain of natural generalized symmetries.
\proclaim Proposition 2.11.
The spinor components,
$$
C_{a}\longleftrightarrow C_{\ss AA^\prime},
$$
of a natural generalized symmetry of order $k$ are functions of the
Penrose fields to order $k$
$$
C_{\ss AA^\prime}=C_{\ss AA^\prime}(\u_{\ss B_0B_1},\ubar_{\ss
B_0^\prime B_1^\prime},\ldots,\u_{\ss B_0\cdots B_k}^{\ss
B_2^\prime\cdots B_{k}^\prime},\ubar^{\ss B_0^\prime \cdots
B_k^\prime}_{\ss B_2\cdots B_{k}}).\tag276
$$

Let us note that the requirements \(222) and \(223) must still be satisfied by
the generalized symmetry \(276).  In particular, the Lorentz invariance
requirement implies that the spinor form of the generalized symmetry must
be $SL(2,{\bf C})$ covariant.  More precisely, if $L^{\ss A}_{\ss B}$ is
an element of $SL(2,{\bf C})$, then
$$
C_{\ss BB^\prime}(L\cdot \u, \overline L\cdot\ubar)
=L^{\ss A}_{\ss B}\overline L^{\ss A^\prime}_{\ss B^\prime}
C_{\ss AA^\prime}(\u,\ubar),\tag277
$$
where $L\cdot\u$ and $\overline L\cdot\ubar$ denote the action of
$SL(2,{\bf C})$ on the Penrose fields, {\it e.g.},
$$
[L\cdot\u]_{\ss AB}=L^{\ss C}_{\ss A}L^{\ss D}_{\ss B}\u_{\ss CD}
\qquad{\rm and}\qquad
[\overline L\cdot\ubar]_{\ss A^\prime B^\prime }=\overline L^{\ss
C^\prime}_{\ss A^\prime}\overline L^{\ss D^\prime}_{\ss
B^\prime}\ubar_{\ss C^\prime D^\prime }.\tag278
$$

To take advantage of Proposition 2.11, we will use the following spinor
form of the linearized equations \(216).

\proclaim Proposition 2.12.
The spinor (and $g$) -valued functions on ${\cal R}^k$
$$
C_{\ss AA^\prime}=C_{\ss AA^\prime}(\u_{\ss B_0B_1},\ubar_{\ss
B_0^\prime B_1^\prime},\ldots,\u_{\ss B_0\cdots B_k}^{\ss
B_2^\prime\cdots B_{k}^\prime},\ubar^{\ss B_0^\prime \cdots
B_k^\prime}_{\ss B_2\cdots B_{k}}).
$$
define a $k^{th}$-order generalized symmetry of the Yang-Mills equations
if and only if
$$
\nabla_{\ss BB^\prime}\nabla^{\ss BB^\prime} C_{\ss AA^\prime}
-\nabla^{\ss BB^\prime}\nabla_{\ss AA^\prime} C_{\ss BB^\prime}
+[\u^{\ss B}_{\ss A},C_{\ss BA^\prime}]
+[\ubar^{\ss B^\prime}_{\ss A^\prime},C_{\ss AB^\prime}]
=0\qquad{\rm on\ }{\cal R}^{k+2}.\tag279
$$

Let us point out that in equation \(279) the covariant derivatives are
defined using total derivatives as in \(217).  In this regard it is worth
noting that the gauge invariance requirement \(223) implies that
$$
\eqalign{
\nabla_{\ss BB^\prime} C_{\ss AA^\prime}
=&{\partial C_{\ss AA^\prime}\over\partial \u_{\ss
C_0C_1}^\alpha}\nabla_{\ss BB^\prime}\u_{\ss C_0C_1}^\alpha
+{\partial C_{\ss AA^\prime}\over\partial \ubar^{\alpha\ss C_0^\prime
C_1^\prime }}\nabla_{\ss BB^\prime}\ubar^{\alpha\ss C_0^\prime
C_1^\prime }\cr
 &+ \cdots +
{\partial C_{\ss AA^\prime}\over\partial \u_{\phantom{\alpha}\ss
C_0C_1\cdots C_k}^{\alpha\ss C_2^\prime\cdots C_k^\prime}}\nabla_{\ss
BB^\prime}\u_{\ss C_0C_1\cdots C_k}^{\alpha\ss C_2^\prime\cdots
C_k^\prime}
+{\partial C_{\ss AA^\prime}\over\partial \ubar^{\alpha\ss C_0^\prime
C_1^\prime \cdots C_k^\prime}_{\phantom{\alpha}\ss C_2\cdots
C_k}}\nabla_{\ss BB^\prime}\ubar^{\alpha\ss C_0^\prime
C_1^\prime\cdots C_k^\prime }_{\phantom{\alpha}\ss C_2\cdots
C_k}.}\tag280
$$

Our analysis of the linearized equation \(279) will involve its
differentiation with respect to the Penrose fields.  Thus we need an
efficient way to deal with symmetric spinors of arbitrary rank.  This will
be done by viewing spinors as multi-linear maps on complex two-
dimensional vector spaces.  If $T^{\ss A_1^\prime\cdots A_q^\prime}_{\ss
A_1\cdots A_p}$ is a spinor of type $(p,q)$ we write
$$
T(\alpha_1,\alpha_2,\ldots,\alpha_p,\alphabar_1,\alphabar_2,\ldots,
\alphabar_q)=T^{\ss A_1^\prime\cdots A_q^\prime}_{\ss A_1\cdots A_p}
\alpha_1^{\ss A_1}\alpha_2^{\ss A_2}\cdots\alpha_p^{\ss A_p}
\alphabar^1_{\ss A_1^\prime}\alphabar^2_{\ss
A_2^\prime}\cdots\alphabar^q_{\ss A_q^\prime}.\tag281
$$
If the spinor $S_{\ss ABC}$ is symmetric in its first two indices, we write
$$
S(\alpha\beta,\gamma)=S_{\ss ABC}\alpha^{\ss
A}\beta^{\ss B}\gamma^{\ss C}=S(\beta\alpha,\gamma),\tag282
$$
where we have dropped the comma between symmetric arguments of $S$.
Note that in this case $S$ is completely determined by the values of
$$
S(\alpha^2,\beta):=S(\alpha\alpha,\beta),\tag283
$$
for all $\alpha$ and $\beta$.
Here we have introduced an exponential notation for repeated symmetric
arguments.  More generally, if $V_{\ss A_1\cdots A_kB}$ is symmetric in
its first $k$ indices, we will write
$$
V(\alpha^k,\beta)=V_{\ss A_1\cdots A_kB}\alpha^{\ss A_1}\cdots
\alpha^{\ss A_k}\beta^{\ss B}.\tag284
$$
We extend our multi-linear map notation to $g$-valued spinors as follows.
If $T=T^\alpha\tau_\alpha$ takes values in the Lie algebra $g$, we will
write
$$
T(v)=T^\alpha v_\alpha,\tag
$$
where $v_\alpha$ are the components of an element $v$ of the dual vector
space $g^*$ to $g$.  If $S$ takes values in $g^*$ we will write
$$
S(w)=S_\alpha w^\alpha,\tag
$$
where $w^\alpha$ are the components of $w\in g$.

The anti-symmetric pairing of spinors defined by the $\epsilon$ spinors is
denoted by
$$
<\alpha,\beta>=\epsilon_{\ss AB}\alpha^{\ss A}\beta^{\ss B}=\alpha_{\ss
A}\beta^{\ss A}\qquad{\rm and}\qquad
<\alphabar,\betabar>=\overline\epsilon_{\ss A^\prime B^\prime
}\alphabar^{\ss A^\prime}\betabar^{\ss B^\prime}=\alphabar_{\ss
A^\prime}\betabar^{\ss A^\prime}.\tag285
$$

Next we develop a notation for derivatives of functions on $J^k(\Q)$ (or
${\cal R}^k$) with respect to the Penrose fields.
If
$$
T{}^{\ss C_1\dots C_p}_{\ss C_1^\prime\dots C_q^\prime}=T{}^{\ss
C_1\dots C_p}_{\ss C_1^\prime\dots C_q^\prime}(\u^1,
\ubar^1,\dots,\u^k,\ubar^k)\tag286
$$
is a natural spinor of type $(p,q)$ and order $k$, then the partial
derivative of
$T{}^{\ss C_1\dots C_p}_{\ss C_1^\prime\dots C_q^\prime}$ with respect
to $\u^l$ is a natural spinor of type $(p+l+1, q+l-1)$.  We shall write
$$
[\partial^l_\u T{}^{\ss C_1\cdots C_p}_{\ss C_1^\prime\dots
C_q^\prime}](\psi^1\cdots\psi^{l+1},\psibar_1\cdots\psibar_{l-1},v)
={\partial T{}^{\ss C_1\dots C_p}_{\ss C_1^\prime\dots
C_q^\prime}\over\partial\u{}^{\alpha\ss A_1^\prime\dots A_{l-
1}^\prime}_{\ss A_1\dots A_{l+1}}}\,
v^\alpha\,\psi^1_{\ss A_1}\cdots\psi^{l+1}_{\ss A_{l+1}}\psibar_1^{\ss
A^\prime_1}\cdots\psibar_{l-1}^{\ss A^\prime_{l-1}}. \tag287
$$

Further, let $\phi^1,\dots,\phi^p$ and $\phibar_1,\dots,\phibar_q$ be
arbitrary spinors of type $(1,0)$ and $(0,1)$ respectively; we shall write
$$
[\partial^l_\u T](\psi^{l+1},\psibar^{l-
1},v;\phi^1,\dots,\phi^p,\phibar_1,\dots,
\phibar_q)=[\partial^l_\u T{}^{\ss C_1\dots C_p}_{\ss C_1^\prime\dots
C_q^\prime}](\psi^{l+1},\psibar^{l-1},v) \phi^1_{\ss
C_1}\cdots\phi^p_{\ss C_p}
\phibar_1^{\ss C^\prime_1}\cdots\phibar_q^{\ss C^\prime_q}.\tag288
$$
A semi-colon will always be used to separate arguments corresponding to
derivatives with respect to the coordinates $\u^k$.
Partial derivatives, $\partial^l_{\ubar}$, with respect to $\ubar_{\ss
A_1\dots A_{l-1}}^{\ss A_1^\prime\dots A_{l+1}^\prime}$ will be
similarly denoted.

We shall repeatedly need certain commutation relations between the partial
derivative operators $\partial_\u$ and $\partial_\ubar$ and the covariant
derivative operator $\nabla^{\ss C}_{\ss C^\prime}$.

\proclaim Lemma 2.13.
Let
$$
T{}^{\cdots}_{\cdots} =
T{}^{\cdots}_{\cdots}(\u^1,\ubar^1,\dots,\u^m,\ubar^m)
$$
be a natural spinor of order $m$.  Then, on ${\cal R}^{m+1}$,
$$
[\partial_\u^{m+1}\nabla^{\ss C}_{\ss C^\prime}T{}^{\cdots}_{\cdots}]
(\psi^{m+2},\psibar\,^{m},v)
=\psi^{\ss C}\psibar\,_{\ss C^\prime}[\partial_\u^mT{}^{\cdots}_{\cdots}]
(\psi^{m+1},\psibar\,^{m-1},v),\tag289
$$
and
$$
[\partial_\u^{m}\nabla^{\ss C}_{\ss C^\prime}T{}^{\cdots}_{\cdots}]
(\psi^{m+1},\psibar\,^{m-1},v)=
[\nabla^{\ss C}_{\ss C^\prime}\partial_\u^{m}T{}^{\cdots}_{\cdots}]
(\psi^{m+1},\psibar\,^{m-1},v)+
\psi^{\ss C}\psibar_{\ss C^\prime}[\partial_\u^{m-1}T^{\cdots}_{\cdots}]
(\psi^{m},\psibar\,^{m-2},v), \tag290
$$
and similarly,
$$
[\partial^{m+1}_\ubar\nabla^{\ss C}_{\ss
C^\prime}T{}^{\cdots}_{\cdots}]
(\psi^{m},\psibar\,^{m+2},v)
=\psi^{\ss C}\psibar_{\ss C^\prime}[\partial^m_\ubar
T{}^{\cdots}_{\cdots}]
(\psi^{m-1},\psibar\,^{m+1},v), \tag291
$$
and
$$
[\partial^{m}_\ubar\nabla^{\ss C}_{\ss C^\prime}T{}^{\cdots}_{\cdots}]
(\psi^{m-1},\psibar\,^{m+1},v)
=[\nabla^{\ss C}_{\ss C^\prime}\partial^{m}_\ubar
T{}^{\cdots}_{\cdots})]
(\psi^{m-1},\psibar\,^{m+1},v)+
\psi^{\ss C}\psibar_{\ss C^\prime}[\partial^{m-1}_\ubar
T^{\cdots}_{\cdots}]
(\psi^{m-2},\psibar\,^{m},v). \tag292
$$

\proof These formulas follow directly from equation \(280) and the
structure equations \(275).

We conclude this section by presenting a couple of elementary results
from spinor algebra which we shall use in our symmetry analysis.
See \refto{CGT1994a} and/or \refto{Penrose1984} for proofs.
\proclaim Lemma 2.14.
Let $P(\psi^k,\alpha)$ be a rank $(k+1)$ spinor that is
symmetric in its first $k$ arguments.  Then there are unique, totally
symmetric spinors $P^*$ and $Q$, of rank $k+1$ and $k-1$
respectively, such that
$$
P(\psi^ k,\alpha)=P^*(\psi^k\alpha)\ + <\psi,\alpha>Q(\psi^{k-1}). \tag293
$$
If $P$ is a natural spinor of the Penrose fields $\u^1$, $\ubar^1$,
\dots,$\u^k$, $\ubar^k$, then so are $P^*$ and $Q$.

\proclaim Lemma 2.15.
Let $P(\psi^k,\alpha)$ be a rank $(k+1)$ spinor that is
symmetric in its first $k$ arguments.  If $P(\psi^k,\alpha)$ satisfies
$$
P(\psi^k,\psi)=0,\tag294
$$
then there is a totally symmetric spinor $Q=Q(\psi^{k-1})$ such that
$$
P(\psi^k,\alpha)=<\psi,\alpha>Q(\psi^{k-1}). \tag295
$$
If $P$ is a natural spinor, then so is $Q$.

\vskip0.25truein
\taghead{3.}
\noindent{\bf 3. Symmetry Analysis.}

We suppose that
$$
C_{\ss AA^\prime}=C_{\ss
AA^\prime}(\u^1,\ubar^1,\ldots,\u^k,\ubar^k)\tag301
$$
is a natural generalized symmetry of order $k$.  Keeping with our
multilinear map notation we write
$$
C(\psi,\psibar,v)=C_{\ss AA^\prime}^\alpha\psi^{\ss A}\psibar^{\ss
A^\prime}v_\alpha.\tag302
$$
On ${\cal R}^{k+2}$ the linearized equation \(279) is a gauge and
$SL(2,{\bf C})$ invariant identity in the Penrose fields $\u^l$ and
$\ubar^l$ for $l=1,\ldots,k+2$.  Our analysis consists of differentiating this
identity with respect to the Penrose fields $\u^l$ and $\ubar^l$ for $l=k,
k+1, k+2$; we present the results in the following series of propositions.
All equations in this section hold on ${\cal R}^{k+2}$, {\it i.e.}, modulo
the field equations.

\proclaim Proposition 3.1.
Let $C_{\ss AA^\prime}=C_{\ss
AA^\prime}(\u^1,\ubar^1,\ldots,\u^k,\ubar^k)$ be a $k^{th}$-order
natural generalized symmetry of the Yang-Mills equations.  Then there
exist natural spinors $G(\psi^k,\psibar^{k-2},v,w)$, $H(\psi^{k-
2},\psibar^k,v,w)$, $A(\psi^k,\psibar^{k},v,w)$,
$B(\psi^{k+2},\psibar^{k-2},v,w)$, $D(\psi^{k-2},\psibar^{k+2},v,w)$,
$E(\psi^k,\psibar^k,v,w)$ such that
$$
\eqalign{
[\partial^k_\u C](\psi^{k+1},\psibar^{k-1},v;\alpha,\alphabar,w)
=&<\alpha,\psi><\alphabar,\psibar>G(\psi^k,\psibar^{k-2},v,w)\cr
&+<\alpha,\psi>A(\psi^k,\psibar^{k-1}\alphabar,v,w)\cr
&+<\alphabar,\psibar>B(\psi^{k+1}\alpha,\psibar^{k-2},v,w),}\tag303
$$
and
$$
\eqalign{
[\partial^k_\ubar C](\psi^{k-1},\psibar^{k+1},v;\alpha,\alphabar,w)
=&<\alpha,\psi><\alphabar,\psibar>H(\psi^{k-2},\psibar^{k},v,w)\cr
&+<\alpha,\psi>D(\psi^{k-2},\psibar^{k+1}\alphabar,v,w)\cr
&+<\alphabar,\psibar>E(\psi^{k-1}\alpha,\psibar^{k},v,w).}\tag304
$$
With the symmetries as indicated, the spinors $A$, $B$, $D$, $E$, $G$,
and $H$ are uniquely determined by $\partial^k_\u C$ and
$\partial^k_\ubar C$.
When $k=1$, equations \(303) and \(304) hold with $B=0$, $D=0$,
$G=0$, and $H=0$.

\proof
This proposition follows from an analysis of the dependence of \(279) on
$\u^{k+2}$ and $\ubar^{k+2}$.  To this end we use the commutation
relations \(289) to find
$$
[\partial^{k+2}_\u \nabla_{\ss MM^\prime}\nabla_{\ss
NN^\prime}C^\alpha_{\ss AA^\prime}](\psi^{k+3},\psi^{k+1},v)
=\psi_{\ss M}\psi_{\ss N}\psibar_{\ss M^\prime}\psibar_{\ss N^\prime}
[\partial^k_\u C^\alpha_{\ss AA^\prime}](\psi^{k+1},\psibar^{k-
1},v).\tag305
$$
We use this result to compute the derivative of \(279) with respect to
$\u^{k+2}$, and this implies that
$$
[\partial^k_\u C](\psi^{k+1},\psibar^{k-1},v;\psi,\psibar,w)=0.\tag306
$$
Similarly, the derivative of \(279) with respect to $\ubar^{k+2}$ implies
that
$$
[\partial^k_\ubar C](\psi^{k-1},\psibar^{k+1},v;\psi,\psibar,w)=0.\tag307
$$
We use Lemmas 2.14 and 2.15 to decompose $[\partial^k_\u C]$ and
$[\partial^k_\ubar C]$ into irreducible components.  We then use \(306)
and \(307) and arrive at \(303) and \(304). Uniqueness of the
decompositions \(303) and \(304) is easily established.\qed

\proclaim Proposition 3.2.
If $C_{\ss AA^\prime}=C_{\ss
AA^\prime}(\u^1,\ubar^1,\ldots,\u^k,\ubar^k)$ is a natural generalized
symmetry of the Yang-Mills equations, then $C_{\ss AA^\prime}$ is linear
in the top-order Penrose fields $\u^k$ and $\ubar^k$.

\proof
This result follows from the quadratic dependence of \(279) on the Penrose
fields $\u^{k+1}$ and $\ubar^{k+1}$.  From Lemma 2.13 we deduce that
$$
\eqalign{
[\partial^{k+1}_\u \partial^{k+1}_\u &\nabla_{\ss MM^\prime}\nabla_{\ss
NN^\prime} C^\alpha_{\ss
AA^\prime}](\chi^{k+2},\chibar^k,u;\psi^{k+2},\psibar^k,v)\cr
&=(\psi_{\ss M}\chi_{\ss N}\psibar_{\ss M^\prime}\chibar_{\ss N^\prime}
+ \psi_{\ss N}\chi_{\ss M}\psibar_{\ss N^\prime}\chibar_{\ss M^\prime})
[\partial^k_\u\partial^k_\u C^\alpha_{\ss AA^\prime}]
(\chi^{k+1},\chibar^{k-1},u;\psi^{k+1},\psibar^{k-1},v).}\tag308
$$
We now differentiate \(279) twice with respect to $\u^{k+1}$ and use
\(308) to find
$$
\eqalign{
2&<\psi,\chi><\psibar,\chibar>[\partial^k_\u\partial^k_\u
C](\chi^{k+1},\chibar^{k-1},u;\psi^{k+1},\psibar^{k-
1},v;\alpha,\alphabar,w)\cr
-&<\psi,\alpha><\psibar,\alphabar>[\partial^k_\u\partial^k_\u
C](\chi^{k+1},\chibar^{k-1},u;\psi^{k+1},\psibar^{k-
1},v;\chi,\chibar,w)\cr
-&<\chi,\alpha><\chibar,\alphabar>[\partial^k_\u\partial^k_\u
C](\psi^{k+1},\psibar^{k-1},v;\chi^{k+1},\chibar^{k-1},u;\psi,\psibar,w)
=0.}\tag309
$$
The last two terms of this equation vanish by virtue of equation \(306) and
we then have
$$
[\partial^k_\u\partial^k_\u C](\chi^{k+1},\chibar^{k-
1},u;\psi^{k+1},\psibar^{k-1},v;\alpha,\alphabar,w)=0.\tag310
$$
Similar computations, which involve applying
$\partial^{k+1}_\u\partial^{k+1}_\ubar$ and
$\partial^{k+1}_\ubar\partial^{k+1}_\ubar$ to the linearized equations,
lead to
$$
[\partial^k_\u\partial^k_\ubar C](\chi^{k+1},\chibar^{k-1},u;\psi^{k-
1},\psibar^{k+1},v;\alpha,\alphabar,w)=0,\tag311
$$
and
$$
[\partial^k_\ubar\partial^k_\ubar C](\chi^{k-1},\chibar^{k+1},u;\psi^{k-
1},\psibar^{k+1},v;\alpha,\alphabar,w)=0.\tag312
$$
\qed

\proclaim Proposition 3.3.
The natural spinors $A$, $B$, $D$, $E$ in the decompositions \(303) and
\(304) depend on the Penrose fields $\u^l$ and $\ubar^l$ for $l\leq k-2$.

\proof
Using Proposition 3.2 it is straightforward to show from the commutation
relations \(289) and \(290) that
$$
\eqalign{
[\partial^k_\u\partial^{k+1}_\u &\nabla_{\ss MM^\prime}\nabla_{\ss
NN^\prime}C^\alpha_{\ss AA^\prime}]
(\chi^{k+1},\chibar^{k-1},u;\psi^{k+2},\psibar^{k},v)\cr
=&\psi_{\ss M}\psi_{\ss N}\psibar_{\ss M^\prime}\psibar_{\ss N^\prime}
[\partial^{k-1}_\u\partial^k_\u C^\alpha_{\ss AA^\prime}]
(\psi^k,\psibar^{k-2},v;\chi^{k+1},\chibar^{k-1},u)\cr
&+(\psi_{\ss M}\chi_{\ss N}\psibar_{\ss M^\prime}\chibar_{\ss N^\prime}
+\chi_{\ss M}\psi_{\ss N}\chibar_{\ss M^\prime}\psibar_{\ss N^\prime})
[\partial^{k-1}_\u\partial^k_\u C^\alpha_{\ss AA^\prime}]
(\chi^k,\chibar^{k-2},u;\psi^{k+1},\psibar^{k-1},v).}\tag313
$$
We now differentiate the linearized equations \(279) with respect to $\u^k$
and $\u^{k+1}$ and use \(313) and \(306) to find
$$
\eqalign{
&2<\psi,\chi><\psibar,\chibar>[\partial^{k-1}_\u\partial^k_\u C]
(\chi^k,\chibar^{k-2},u;\psi^{k+1},\psibar^{k-1},v;\alpha,\alphabar,w)\cr
&-<\psi,\alpha><\psibar,\alphabar>[\partial^{k-1}_\u\partial^k_\u C]
(\psi^k,\psibar^{k-2},v;\chi^{k+1},\chibar^{k-1},u;\psi,\psibar,w)\cr
&-<\psi,\alpha><\psibar,\alphabar>[\partial^{k-1}_\u\partial^k_\u C]
(\chi^k,\chibar^{k-2},u;\psi^{k+1},\psibar^{k-
1},v;\chi,\chibar,w)=0.}\tag314
$$
We set $\alpha=\psi$ and find
$$
[\partial^{k-1}_\u\partial^k_\u C]
(\chi^k,\chibar^{k-2},u;\psi^{k+1},\psibar^{k-
1},v;\psi,\alphabar,w)=0.\tag315
$$
In terms of the decomposition \(303), this equation implies that
$$
[\partial^{k-1}_\u B](\chi^k,\chibar^{k-2},u;\psi^{k+2},\psibar^{k-2},v,w)
=0,\tag316
$$
{\it i.e.}, $B$ is independent of the Penrose fields $\u^{k-1}$.  In a similar
fashion, setting $\alphabar=\psibar$ leads to
$$
[\partial^{k-1}_\u A](\chi^k,\chibar^{k-
2},u;\psi^k,\psibar^k,v,w)=0.\tag317
$$
Analogous computations, which involve applying the derivatives
$\partial_\u^k\,\partial_\ubar^{k+1}$,
$\partial_\ubar^k\,\partial_\u^{k+1}$, and
$\partial_\ubar^k\,\partial_\ubar^{k+1}$ to \(279), yield
$$
\eqalign{
&[\partial^{k-1}_\u D](\chi^k,\chibar^{k-2},u;\psi^{k-
2},\psibar^{k+2},v,w)=0,\cr
&[\partial^{k-1}_\u E](\chi^k,\chibar^{k-
2},u;\psi^k,\psibar^k,v,w)=0,}\tag318
$$
and
$$
\eqalign{
&[\partial^{k-1}_\ubar A](\chi^{k-
2},\chibar^k,u;\psi^k,\psibar^k,v,w)=0,\cr
&[\partial^{k-1}_\ubar B](\chi^{k-2},\chibar^k,u;\psi^{k+2},\psibar^{k-
2},v,w)
=0,\cr
&[\partial^{k-1}_\ubar D](\chi^{k-2},\chibar^k,u;\psi^{k-
2},\psibar^{k+2},v,w)=0,\cr
&[\partial^{k-1}_\ubar E](\chi^{k-
2},\chibar^{k},u;\psi^k,\psibar^k,v,w)=0.}\tag318a
$$
\qed

\proclaim Proposition 3.4.
Let $C_{\ss AA^\prime}$ be a natural generalized symmetry of order
$k>1$.  Then there is a natural g-valued function of order $k-1$,
$$
\Lambda=\Lambda(\u^1,\ubar^1,\ldots,\u^{k-1},\ubar^{k-1}),
$$
such that, in the decompositions \(303) and \(304) for $C_{\ss
AA^\prime}$, $G$ and $H$ are the gradients
$$
\eqalign{
G(\psi^k,\psibar^{k-2},v,w)&=[\partial^{k-1}_\u \Lambda]
(\psi^k,\psibar^{k-2},v;w)\cr
H(\psi^{k-2},\psibar^{k},v,w)&=[\partial^{k-1}_\ubar\Lambda]
(\psi^{k-2},\psibar^{k},v;w).}\tag319
$$

\proof
We begin by deriving the integrability conditions for \(319) from the
linearized equations \(279).  We return to equation \(314), which, on
account of Propositions 3.1 and 3.3, reduces to
$$
[\partial^{k-1}_\u G](\chi^k,\chibar^{k-2},u;\psi^k,\psibar^{k-2},v,w)
=[\partial^{k-1}_\u G](\psi^k,\psibar^{k-2},v;\chi^k,\chibar^{k-
2},u,w).\tag321
$$
This is one of the integrability conditions needed to establish \(319).  The
remaining integrability conditions,
$$
[\partial^{k-1}_\ubar H](\chi^{k-2},\chibar^{k},u;\psi^{k-
2},\psibar^{k},v,w)
=[\partial^{k-1}_\ubar H](\psi^{k-2},\psibar^{k},v;\chi^{k-
2},\chibar^{k},u,w),\tag3211
$$
and
$$
[\partial^{k-1}_\ubar G](\chi^{k-2},\chibar^{k},u;\psi^k,\psibar^{k-
2},v,w)
=[\partial^{k-1}_\u H](\psi^{k},\psibar^{k-2},v;\chi^{k-
2},\chibar^{k},u,w)\tag3212
$$
are obtained in an analogous manner from the equations resulting from
applying $\partial_\u^k\,\partial_\ubar^{k+1}$,
$\partial_\ubar^k\,\partial_\u^{k+1}$, and
$\partial_\ubar^k\,\partial_\ubar^{k+1}$ to \(279).

{}From these integrability conditions it is straightforward to verify that
$\Lambda$ can be expressed as a natural function of order $k-1$ via
$$
\eqalign{
\Lambda^\alpha&=\int_0^1dt\, \u^{\beta\ss A_1\cdots A_k}_{\ss
A_1^\prime\cdots A_{k-2}^\prime} G_{\beta\ss A_1\cdots A_k}^{\alpha\ss
A_1^\prime\cdots A_{k-2}^\prime}(\u^1,\ubar^1,\ldots,t\u^{k-
1},t\ubar^{k-1})\cr
&+\int_0^1dt\, \ubar^{\beta\ss A_1\cdots A_{k-2}}_{\ss A_1^\prime\cdots
A_{k}^\prime} H_{\beta\ss A_1\cdots A_{k-2}}^{\alpha\ss
A_1^\prime\cdots A_{k}^\prime}(\u^1,\ubar^1,\ldots,t\u^{k-1},t\ubar^{k-
1}).}\tag
$$
\qed

\proclaim Proposition 3.5.
Let $C_{\ss AA^\prime}$ be a natural generalized symmetry of order $k$.
Then there is a natural $g$-valued function
$\Lambda=\Lambda(\u^1,\ubar^1,\ldots,\u^{k-1},\ubar^{k-1})$ and a
natural generalized symmetry $\widehat  C_{\ss AA^\prime}$ of order $k-
1$ such that
$$
C_{\ss AA^\prime}=\widehat  C_{\ss AA^\prime}+\nabla_{\ss
AA^\prime}\Lambda.\tag322
$$

\proof
We choose $\Lambda$ as in Proposition 3.4 and define
$$
\widehat C_{\ss AA^\prime}=C_{\ss AA^\prime}-\nabla_{\ss
AA^\prime}\Lambda.\tag323
$$
By Proposition 2.4 and linearity of the equations \(279), $\widehat C_{\ss
AA^\prime}$ is a generalized symmetry; by construction,
$\partial^k_\u\widehat C_{\ss AA^\prime}$ has the decomposition
$$
\eqalign{
[\partial^k_\u \widehat C](\psi^{k+1},\psibar^{k-1},v;\alpha,\alphabar,w)
=&<\alpha,\psi>A(\psi^k,\psibar^{k-1}\alpha,v,w)\cr
&+<\alphabar,\psibar>B(\psi^{k+1}\alpha,\psibar^{k-2},v,w),}\tag324
$$
and
$\partial^k_\ubar\widehat C_{\ss AA^\prime}$ has the decomposition
$$
\eqalign{
[\partial^k_\ubar \widehat C](\psi^{k-
1},\psibar^{k+1},v;\alpha,\alphabar,w)
=&<\alpha,\psi>D(\psi^{k-2},\psibar^{k+1}\alpha,v,w)\cr
&+<\alphabar,\psibar>E(\psi^{k-1}\alpha,\psibar^{k},v,w).}\tag325
$$

We now show that the linearized equations \(279) force $A$, $B$, $D$,
and $E$ to vanish, thus establishing \(322).   To this end, we consider the
derivative of the linearized equations \(279) with respect to $\u^{k+1}$.
We use the commutation relation \(290) and equation \(306) to find that
$$
\eqalign{
&2\psi^{\ss A}\psibar^{\ss A^\prime}\nabla_{\ss AA^\prime}
[\partial^k_\u \widehat C](\psi^{k+1},\psibar^{k-
1},v;\alpha,\alphabar,w)\cr
&-<\alpha,\psi><\alphabar,\psibar>w_\alpha\nabla^{\ss AA^\prime}
[\partial^k_\u \widehat C^\alpha_{\ss AA^\prime}](\psi^{k+1},\psibar^{k-
1},v)\cr
&-<\alpha,\psi><\alphabar,\psibar>[\partial^{k-1}_\u \widehat C]
(\psi^k,\psibar^{k-2},v;\psi,\psibar,w)=0.}\tag326
$$
In this equation we set $\alpha=\psi$ and substitute from \(303) to obtain
$$
\psi^{\ss A}\psibar^{\ss A^\prime}[\nabla_{\ss
AA^\prime}B](\psi^{k+2},\psibar^{k-2},v,w)=0.\tag327
$$
Similarly, setting $\alphabar=\psibar$ we obtain
$$
\psi^{\ss A}\psibar^{\ss A^\prime}[\nabla_{\ss
AA^\prime}A](\psi^{k},\psibar^{k},v,w)=0.\tag328
$$
These equations imply that $A$ and $B$ are independent of the Penrose
fields $\u^l$ and $\ubar^l$, for $l=1,\dots,k-2$.  To see this, let us consider
the spinor $A$.  If we assume $A$ is a natural spinor of order $l$, then
the derivative of \(328) with respect to $\u^{l+1}$ becomes, after using the
commutation relation \(289),
$$
<\chi,\psi><\chibar,\psibar>[\partial^l_\u A](\chi^{l+1},\chibar^{l-
1},u;\psi^k,\psibar^k,v,w)=0,\tag329
$$
which implies
$$
\partial^l_\u A = 0.\tag330
$$
A simple induction argument then shows that $A$ is independent of all the
Penrose fields $\u^l$  for $l=1,\dots,k-2$.  An identical argument
establishes that $A$ is independent of $\ubar^l$  for $l=1,\dots,k-2$.  In a
similar fashion we can show that $B$ is independent of the Penrose fields
$\u^l$ and $\ubar^l$, for $l=1,\dots,k-2$.  We conclude that $A$ and $B$
are $SL(2,{\bf C})$ invariant spinors constructed solely from the
$\epsilon$ spinors.  But there are no $SL(2,{\bf C})$ invariant spinors
with the rank and symmetry of $A$ or $B$ built solely from the $\epsilon$
spinors, so $A$ and $B$ must vanish.

If we differentiate the linearized equations for $\widehat C$ with respect to
$\ubar^{k+1}$, a similar line of reasoning shows that $D$ and $E$ must
also vanish.\qed

We can now classify all natural generalized symmetries of the Yang-Mills
equations.

\proclaim Theorem 3.6.
Let
$$
C_a = C_a(x, A_a,A_{a,b_1}, \ldots,A_{a,b_1\cdots b_k})
$$
be a $k^{th}$-order natural generalized symmetry of the Yang-Mills
equations.  Then
there is a natural $g$-valued function
$$
\Lambda=\Lambda(x, A_a,A_{a,b_1}, \ldots,A_{a,b_1\cdots b_{k-1}}),
$$
such that, modulo the field equations,
$$
C_a=\nabla_a\Lambda.
$$

\proof
{}From Proposition 3.5 we have that every generalized symmetry of order
$k$ differs from a symmetry of order $k-1$ by a generalized gauge
symmetry.  By induction, every generalized symmetry of order $k$ differs
from a gauge symmetry by a generalized symmetry of order $1$, which
we denote by $C^{(1)}_{\ss AA^\prime}$.  From Proposition 3.1 and 3.3,
we can apply equation \(326) to $C^{(1)}_{\ss AA^\prime}$.  From the
discussion following \(326) we conclude that $C^{(1)}_{\ss AA^\prime}$
is in fact independent of the Penrose fields and is thus an $SL(2,{\bf C})$
invariant spinor of type $(1,1)$ constructed from the $\epsilon$ spinors.
But there are no such spinors, as can be seen, for example, by noting that
such a spinor would define a Lorentz invariant vector field.  And so it
follows that $C^{(1)}_{\ss AA^\prime}=0$.\qed
\vskip0.25truein
\noindent{\bf 4. Discussion.}

We have shown that all natural generalized symmetries of the Yang-Mills
equations are generalized gauge symmetries.  These symmetries are
physically trivial, and they give rise to trivial conservation laws.   In order
to extend our results to a complete classification of
generalized symmetries of the Yang-Mills equations we will have to drop
the requirements \(222) and \(223).  Thus we must consider solutions of the
linearized equations \(216) which are (i) not gauge covariant, and (ii) not
Poincar\' e covariant, {\it i.e.}, $C_a$ is now allowed to be any function of
the coordinates \(229) or, better yet, the coordinates \(273).  In the
gravitational case \refto{IMA_CGT}, the generalizations analogous to (i)
and (ii) lead to no
new types of symmetries.  Preliminary computations imply that (i) is
unlikely to lead to any new symmetries also in the Yang-Mills case for
similar reasons to those found in \refto{IMA_CGT}.  On the other hand,
the relaxation of
Poincar\' e invariance may lead to new, non-trivial symmetries (beyond
those of Proposition 2.2).  Indeed, the putative generalized symmetries can
be
constructed using the conformal Killing vectors admitted by the underlying
Minkowski spacetime, and this significantly changes the analysis beginning
with Proposition 3.5.  We will present the complete symmetry analysis
elsewhere.
\vskip 0.5truein
\noindent{\bf Acknowledgment}

I would like to thank Ian Anderson for helpful discussions.
\vfill\eject
\references

\refis{footnote1}{While the infinitesimal transformations are local in the
fields and their derivatives, the corresponding finite transformations need
not be local.}

\refis{IMA_CGT}{C. G. Torre  and I. M. Anderson, \prl 70, 3525, 1993;
I. M. Anderson and C. G. Torre, ``Classification of generalized symmetries
for the vacuum Einstein equations'', Utah State University preprint, 1994.}

\refis{Cheng1984}{T. Cheng and L. Li, {\it Gauge Theory of Elementary
Particle Physics}, (Oxford University Press, Oxford 1984).}

\refis{Ablowitz1993}{M. Ablowitz, S. Chakravarty, and L. Takhtajan,
\cmp 158, 289, 1993.}

\refis{Olver1993}{P. Olver, {\it Applications of Lie Groups to Differential
Equations}, (Springer-Verlag, New York 1993).}

\refis{Bluman1989}{G. Bluman and S. Kumei, {\it Symmetries of Differential
Equations}, (Springer-Verlag, New York 1989).}

\refis{Fokas1987}{A. Fokas, \journal Stud. Appl. Math., 77, 253, 1987.}

\refis{Mikhailov1991}{A. Mikhailov, A. Shabat, and V. Sokolov in {\it What
is Integrability?}, ed. V. Zakharov (Springer-Verlag, New York 1991).}

\refis{Penrose1960}{R. Penrose, \ann 10, 171, 1960.}

\refis{Penrose1984}{R. Penrose and W. Rindler, {\it Spinors and Space-Time,
Vol. 1}, (Cambridge University Press, Cambridge 1984).}

\refis{Saunders1989}{D. Saunders, {\it The Geometry of Jet Bundles},
(Cambridge University Press, Cambridge 1989).}

\refis{CGT1994a}{C. G. Torre and I. M. Anderson, {\it Two-Component Spinors
and Natural Coordinates for the Prolonged Einstein Equation Manifolds},
{\sl Utah State University Technical Report}, 1994.}

\endreferences
\bye